\documentclass[pre,aps,twocolumn,showpacs]{revtex4}
\usepackage{graphicx}
\usepackage{dcolumn}
\usepackage{bm}
\usepackage{color}
\usepackage{verbatim}
\usepackage{epstopdf}
\usepackage{amsthm}
\usepackage{natbib}
\usepackage{hyperref}
\usepackage{amsmath}
\usepackage{amsfonts}
\usepackage{amssymb}%
\providecommand{\U}[1]{\protect\rule{.1in}{.1in}}

\theoremstyle{definition}

\theoremstyle{remark}

\begin{document}
\title{Numerical Computations of Separability Probabilities}
\author{Jianjia Fei}
\email{jfei@wisc.edu}
\affiliation{Department of Physics, University of Wisconsin-Madison, Madison, Wisconsin 53706, USA}
\author{Robert Joynt}
\email{rjjoynt@wisc.edu}
\affiliation{Department of Physics, University of Wisconsin-Madison, Madison, Wisconsin 53706, USA}
\date{\today}

\begin{abstract}
We compute the probability that a bipartite quantum state is separable by
Monte Carlo sampling. 
This is carried out for rebits, qubits and quaterbits.
We sampled $5\times10^{11}$ points for each of these three cases.
The results strongly support conjectures for certain rational values of
these probabilities that have been found by other methods. 

\end{abstract}

\maketitle

\section{Introduction}

The surprising efficacy of complex numbers in descrbing the physical world has
led to persistent speculation that quaternions might also serve as a fruitful
foundation of physical theories. 
Quaternions resemble the complex numbers in forming a division ring: they are 
the richest such number system that has the very restrictive unique division 
property (Frobenius). 
Thus there is a natural mathematical progression from the real to the complex 
to the quaternionic numbers.
We might ask if there is a corresponding natural progression also in physical 
theories that use these numbers. \ 

Real numbers are sufficient to completely describe rebits, physical objects
with two degrees of freedom whose $2\times2$ density matrices are symmetric
real matrices. 
Complex numbers describe the usual qubits that have $2\times2$ complex 
Hermitian density matrices. 
Quaterbits are described by $2\times2$ density matrices with quaternionic 
entries. 
These matrices are Hermitian in the sense that the transpose is the 
quaternionic conjugate. 
The conjugate of a quaternion $h=a+i_{h}b+jc+kd$ is 
$\overline{h}=a-i_{h}b-jc-kd.$
$a,b,c,d$ are real and Hamilton's symbols $i_{h},j,k$ satisfy 
$i_{h}^{2}=j^{2}=k^{2}=-1,i_{h}j=-ji_{h}=k,$ etc. 
The density matrices are positive and have unit trace in all cases.\ \ \ 

Recently, there has been some interesting mathematical work that gives some
indication of such a natural progression in the properties of these density
matrices. 
This progression arises in the context of considering correlations
in bipartite systems, i.e., in the $4\times4$ density matrices $\rho$ that
describe a pair of physical objects. 
Two objects $A$ and $B$ are said to be separable if $\rho$ may be written as
\begin{equation}
\rho=\sum_{i}p_{i}~\rho_{i}^{A}\otimes\rho_{i}^{B}
\end{equation}
where the real numbers $p_{i}$ satisfy $p_{i}\geq0$ and $\sum_{i}p_{i}=1.$
\ $\rho_{i}^{A}$ and $\rho_{i}^{B}$ are $2\times2$ density matrices that refer
to the 2 objects individually. 
Clearly this definition makes sense in all three number systems, as do the 
positivity and trace conditions. 

Now there is a very fundamental question, first proposed in~\cite{Zyczkowski1998}: 
what proportion $P$ of bipartite systems are separable? There is an intriguing
conjecture that the formula
\begin{equation}
P\left( \alpha\right) = \sum_{i=0}^{\infty} f(\alpha + i),
\label{eq:conj}
\end{equation}
where 
\begin{equation}
f\left(  \alpha\right)  =
\frac{q\left(  \alpha\right)  2^{-4\alpha-6}\Gamma\left(  3\alpha+\frac{5}
{2}\right)  \Gamma\left(  5\alpha+2\right)  }{3\Gamma\left(  \alpha+1\right)
\Gamma\left(  2\alpha+3\right)  \Gamma\left(  5\alpha+\frac{13}{2}\right)
},
\end{equation}
with
\begin{align}
q\left(  \alpha\right) & = 185000\alpha^{5}+779750\alpha^{4}+1289125\alpha^{3}\notag \\
                       & +1042015\alpha^{2}+410694\alpha+63000,
\end{align}
which for simple integral and half-odd-integral values of $\alpha$ gives
results that are close to rational numbers with fairly small denominators, and
that the rational numbers $P\left(  1/2\right)  $ gives the proportion of
separable 2-rebit states, and $P\left(  1\right)  $ gives the proportion of
separable 2-qubit states. 
This remarkable result~\cite{Slater2012,Slater2013} comes from the
computation of moments such as $\left\langle \left\vert \rho\right\vert
^{n}\ \left\vert \rho^{PT}\right\vert \right\rangle ,$ an application of
Zeilberger's algorithm~\cite{Hou2011}, and numerical evaluation of 
Eq. \ref{eq:conj} to thousands of decimal places. 
Here the angle brackets refer to an average over all physical states (i.e., 
those that satisfy positivity and have unit trace), and the straight brackets 
denote the determinant. 
The average is defined using the measure induced by the Hilbert-Schmidt metric 
on the space of density matrices. 
As we shall see, the $\alpha=1$ and $\alpha=1/2$ conjectures are extremely 
well-supported. 
So it is natural to ask if higher values of $\alpha$ have a physical 
interpretation. 
In this paper, we shall focus on the possibility that $\alpha=2$ corresponds 
to quaterbits (quaternion-based bits), though we also present computations for 
qubits and rebits. 
Our approach is to compute $P\left(  \alpha\right)  $ by a Monte Carlo method.

\section{Qubits}

For qubits, the conjecture based on Eq. \ref{eq:conj} gives
\[
P\left(  1\right)  =\frac{8}{33}=0.\overline{24},
\]
where the overbar indicates a repeating decimal. \ 

This conjecture was also supported by numerical evidence~\cite{Zhou2012}, 
with Monte Carlo simulations yielding
\[
P_{est}\left(  1\right)  = 0.2424 \pm 0.0002.
\]
This earlier Monte Carlo work used the condition of zero concurrence for
separabilty~\cite{Hill1997, Wooters1998}. 
This is equivalent to the Peres-Horodecki Criterion (PHC)
~\cite{Peres1996,Horodecki1996} which states that to be separable, it is 
necessary and sufficient for the partial transpose $\rho^{PT}$ of the 
density matrix $\rho$ to be positive. 
In the 2-qubit case, the $4\times4$ density matrix can be written as
\begin{equation}
\rho=\frac{1}{4}I_{4}+\frac{1}{4}\sum_{i,j=0}^{3}n_{ij}\left(  \sigma_{i}\otimes
\sigma_{j}\right)  ,\label{eq:rho}
\end{equation}
where $I_{n}$ is the $n\times n$ identity matrix, $\sigma_{0}=I_{2},$ and the
$\sigma_{i}$ are the Pauli matrices. The sum excludes the $\,i=j=0$ term, and
$n_{ij}$ is thus a real 15-vector. 
We can use Euclidean measure to define probabilities in the space, which is 
equivalent to the Hilbert-Schmidt measure. The allowed values of $n_{ij}$ form 
a compact and convex subset of R$^{15}$ whose boundary is set by the condition 
that $\rho$ is positive$.$ 
This set is the generalization of the familiar Bloch sphere for spin 1/2.
Its shape has been described in at least a partial fashion~\cite{Zhou2012,Byrd2003}, 
and its volume has been computed~\cite{Andai2006}. 
It lies within the sphere given by
$\sum_{i,j=0}^{3}n_{ij}^{2}=3/4.$ 
The 15-ball of radius $\sqrt{3/4}$ is sampled uniformly, testing both positivity 
and the PHC, which yields $P\left(1\right)$. 
The sampling method is taken from~\cite{Watson1983}. 
We sampled $5\times10^{11}$ points, obtaining
\[
P_{est}\left(  1\right)  = 0.2424 \pm 0.0001.
\]
Thus the numerical results strongly support the conjecture $P\left(  1\right)
=8/33.$ \ 

\section{Rebits}

The progression aspect of the problem arises already when we consider the same
problem for rebits. 
Rebits are obtained by setting $n_{02}=n_{12}=n_{32}=n_{20}=n_{21}=n_{23}=0$, 
\textit{i.e.}, omitting the imaginary generators in Eq. \ref{eq:rho}$.$ 
Positivity still requires that $\sum n_{ij}^{2}\leq3/4$ for the coefficients 
of the nonzero generators. 
The conjecture is that
\[
P\left(  1/2\right)  = \frac{29}{64} = 0.453125.
\]
What is remarkable is that $P\left(  1\right)  $ and $P\left(  \frac{1}
{2}\right)  $ are given by the same formula, changing only the parameter
$\alpha.$ 
We have performed the Monte Carlo sampling for this case, testing positivity 
and the PHC for points in the 9-ball. 
The result is
\begin{align*}
P\left(  1/2\right) &  = 0.4531 \pm 0.0001, ~\text{while}\\
\frac{29}{64} & = 0.453125.
\end{align*}
We tested $5\times10^{11}$ points.
Hence the unified formula is well-confirmed by the numerical computations for 
both $\alpha=1/2$ and $\alpha=1$.

\section{Quaterbits}

Now we consider quaterbits. 
It is reasonable to conjecture that $\alpha=2$ formula should give the 
separability ratio for this 26-dimensional case. 
The conjecture is:
\[
P\left(  2\right)  =\frac{26}{323}\approx0.080495.
\]
We first note that the PHC has not been proven for this case - it is not known
whether positivity of the partial transpose is equivalent to separability.
Thus it is very interesting to repeat the above calculations for this case.
The $2\times2$ matrix representation of quaternions in which $h=a+i_{h}
b+ic+id\rightarrow I_{2}a+ib\sigma_{x}+ic\sigma_{y}+id\sigma_{z}$ will be
useful. 
$I_{n}$ is the $n\times n$ identity.

Two quaterbits are described by $4\times4$ density matrices $\rho$ with
quaternionic entries. 
These matrices are self adjoint. 
Writing the quaternions themselves as matrices we find

\[
\rho-\frac{I_{8}}{8}=\rho^{\prime}=
\begin{pmatrix}
AI_{2} & q_{0} & q_{1} & q_{2}\\
\overline{q}_{0} & BI_{2} & q_{3} & q_{4}\\
\overline{q}_{1} & \overline{q}_{3} & CI_{2} & q_{5}\\
\overline{q}_{2} & \overline{q}_{4} & \overline{q}_{5} & DI_{2}
\end{pmatrix}
,
\]
with
\[
q_{i}=
\begin{pmatrix}
a_{i}-id_{i} & ib_{i}+c_{i}\\
ib_{i}-c_{i} & a_{i}+id_{i}
\end{pmatrix}
,~~\overline{q}_{i}=
\begin{pmatrix}
a_{i}+id_{i} & -ib_{i}-c_{i}\\
-ib_{i}+c_{i} & a_{i}-id_{i}
\end{pmatrix}
.
\]
We must have that $A+B+C+D=0.$ 
Defining $u=2A+2B,v=2A+2C,w=-2B-2C$ and
\[
\lambda_{ijk}=\sigma_{i}\otimes\sigma_{j}\otimes\sigma_{k}
\]
so that $Tr$ $\lambda_{ijk}\lambda_{i^{\prime}j^{\prime}k^{\prime}}
=8~\delta_{ii^{\prime}}\delta_{jj^{\prime}}\delta_{kk^{\prime}},$ we find,
after a lengthy calculation:
\begin{align*}
\rho^{\prime}  & =u\lambda_{300}+v\lambda_{030}+w\lambda_{330} \\
& +\frac{1}{2}a_{0}\left(  \lambda_{010}+\lambda_{310}\right)  +\frac{1}{2}b_{0}\left(  \lambda_{021}+\lambda_{321}\right)  -\frac{1}{2}c_{0}\left(  \lambda_{022}+\lambda_{322}\right) \\
& -\frac{1}{2}d_{0}\left(  \lambda_{023}+\lambda_{323}\right)  +\frac{1}{2}a_{1}\left(  \lambda_{100}+\lambda_{130}\right)  -\frac{1}{2}b_{1}\left(  \lambda_{201}+\lambda_{231}\right) \\
& -\frac{1}{2}c_{1}\left(  \lambda_{202}+\lambda_{232}\right)  -\frac{1}{2}d_{1}\left(  \lambda_{203}+\lambda_{233}\right)  +\frac{1}{2}a_{2}\left(  \lambda_{110}-\lambda_{220}\right) \\
& -\frac{1}{2}b_{2}\left(  \lambda_{121}+\lambda_{211}\right)  -\frac{1}{2}c_{2}\left(  \lambda_{122}+\lambda_{212}\right)  -\frac{1}{2}d_{2}\left(  \lambda_{123}+\lambda_{213}\right) \\
& +\frac{1}{2}a_{3}\left(  \lambda_{110}+\lambda_{220}\right)  +\frac{1}{2}b_{3}\left(  \lambda_{121}-\lambda_{211}\right)  -\frac{1}{2}c_{3}\left(  \lambda_{122}-\lambda_{212}\right) \\
& +\frac{1}{2}d_{3}\left(  \lambda_{123}-\lambda_{213}\right)  +\frac{1}{2}a_{4}\left(  \lambda_{100}-\lambda_{130}\right)  -\frac{1}{2}b_{4}\left(  \lambda_{201}-\lambda_{231}\right) \\
& +\frac{1}{2}c_{4}\left(  \lambda_{202}-\lambda_{232}\right)  -\frac{1}{2}d_{4}\left(  \lambda_{023}-\lambda_{233}\right)  +\frac{1}{2}a_{5}\left(  \lambda_{010}-\lambda_{310}\right) \\
& -\frac{1}{2}b_{5}\left(  \lambda_{021}-\lambda_{321}\right)  +\frac{1}{2}c_{5}\left(  \lambda_{022}-\lambda_{322}\right)  -\frac{1}{2}d_{5}\left(  \lambda_{023}-\lambda_{323}\right)
\end{align*}
Thus%
\[
\rho^{\prime}=\sum_{ijk}^{\prime}n_{ijk}\lambda_{ijk},
\]
where the sum $\Sigma^{^{\prime}}$ runs only over the combinations%
\begin{align*}
\left\{  ijk\right\} = &\left\{  300\right\}  ,\left\{  030\right\}  ,\left\{  330\right\}  ,\left\{  010\right\}  ,\left\{  310\right\}  ,\left\{  021\right\}  ,\\
                       &\left\{  321\right\}  ,\left\{  022\right\}  ,\left\{  322\right\}  ,\left\{  023\right\}  ,\left\{  323\right\}  ,\left\{  100\right\}  ,\\
                       &\left\{  130\right\}  ,\left\{  201\right\}  ,\left\{  231\right\}  ,\left\{  202\right\}  ,\left\{  232\right\}  ,\left\{  203\right\}  ,\\
                       &\left\{  233\right\}  ,\left\{  110\right\}  ,\left\{  220\right\}  ,\left\{  121\right\}  ,\left\{  211\right\}  ,\left\{  122\right\}  ,\\
                       &\left\{  212\right\}  ,\left\{  123\right\}  ,\left\{  213\right\}  .
\end{align*}
and positivity requires that
\[
\sum_{ijk}^{\prime}\left(  n_{ijk}\right)  ^{2}\leq\frac{7}{64}.
\]

To determine $P\left(  2\right)  $ numerically, we sample the 27-ball of
radius $\sqrt{7/64}$ uniformly in the $n_{ijk},$ which, as stated above, is
also uniform in the Hilbert-Schmidt metric. 
We test each point for PPT and positivity, giving an estimate 
$P_{est}\left(  2\right)  .$
$5\times 10^{11}$ points are sampled in the Monte Carlo simulation.
We find
\[
P_{est}\left(  2\right)  = 0.0805 \pm 0.0001.
\]
The numerical results give strong evidence in favor of the conjecture.

\section{Conclusion}

Quaternionic quantum mechanics has been investigated in detail. 
It can only describe the observed universe if some superselection rules are 
added~\cite{Adler1995}.
Rebits do not have a rich enough mathematical structure to describe the 
real world - it would be very difficult to see how a rebit could display Ramsey 
fringes, for example, since the whole Bloch sphere is required for the dynamics. 
Qubits seem to be about right, of course.
But it is remarkable that some mathematical structures overarch the three 
possibilities. 
The separability probability formula in Eq. \ref{eq:conj} seems to be one of 
these.\bigskip



\begin{acknowledgments}

We thank Dong Zhou for useful discussions. We thank P. B. Slater and K. Zyczkowski for helpful communications. We also thank the HEP, Condor, and CHTC groups at UW-Madison for computational support.

\end{acknowledgments}



\end{document}